\documentclass[%
 reprint,
superscriptaddress,
 amsfonts,amsmath,amssymb,amsthm,
 aps,
prx,
]{revtex4-1}

\usepackage{amsmath}
\usepackage{graphicx}
\usepackage{dcolumn}
\usepackage{natbib}
\usepackage{comment}
\usepackage[dvipsnames]{xcolor}

\begin{document}

\title{Multi-Gbps quantum randomness source based on direct detection and vacuum states}

\author{Dino Solar Nikolic}
\email{These authors contributed equally.}
\affiliation{Center for Macroscopic Quantum States (bigQ), Department of Physics, Technical University of Denmark, 2800 Kongens Lyngby, Denmark}

\author{Cosmo Lupo}
\email{These authors contributed equally.}
\affiliation{Dipartimento Interateneo di Fisica, Politecnico \& Universit\`a di Bari, 70126 Bari, Italy}
\affiliation{INFN, Sezione di Bari, 70126 Bari, Italy}

\author{Runjia Zhang}
\email{These authors contributed equally.}

\author{Tobias Rydberg}
\author{Ulrik L. Andersen}
\author{Tobias Gehring}
\email{tobias.gehring@fysik.dtu.dk}
\affiliation{Center for Macroscopic Quantum States (bigQ), Department of Physics, Technical University of Denmark, 2800 Kongens Lyngby, Denmark}

\date{\today}

\begin{abstract}
Quantum random number generators (QRNGs) based on quadrature measurements of the vacuum have so far used balanced homodyne detection to obtain a source of high entropy. Here we propose a simple direct detection measurement scheme using only a laser and a photodiode that still extracts randomness from vacuum fluctuations. We prove the security of the QRNG based on a reduced set of assumptions in comparison to previous security proofs for quadrature detection as our proof does not require the laser or electronic noise to be Gaussian.
Using a low-cost setup based on a vertical-cavity surface-emitting laser we experimentally implement the QRNG scheme. We propose a system characterization method, apply it to our implementation and demonstrate a real-time randomness extraction rate of 3.41 Gbit per second.
The unique combination of speed, low cost, and rigorous security proof gives our QRNG design a large potential for a wide-scale usage in a variety of applications ranging from quantum key distribution to mobile applications and internet of things.
\end{abstract}

\maketitle

\section{Introduction}

Quantum random number generators (QRNGs) use entropy from measurements on quantum states to produce random numbers of highest quality~\cite{Ma2016,Acin2016,Herrero-collantes2017}. 
The laws of physics then guarantee that the random numbers are truly unpredictable.
QRNGs can generate random numbers at rates in the Gbit per second regime which makes them suitable for applications of highest demand on security and speed, for instance quantum key distribution~\cite{Pirandola2020}, as well as conventional cybersecurity.

Many different physical systems have been explored as QRNGs. Among them, single photons impinging on a beam splitter~\cite{Jennewein2000,Stafanov2000}, photon counting~\cite{Ren2011QuantumDetector}, measurements of phase noise~\cite{Nie2015}, measurements of amplified spontaneous emission of lasers~\cite{Abellan2014,Martin2014}, and quadrature measurements on vacuum states~\cite{Gabriel2010,Haw2015,Shi2016,Zhang2016,Gehring2021Homodyne-basedSide-information,Bruynsteen2022}. In particular the latter two approaches can reach the high speed regime required by quantum key distribution with generation speeds in the excess of 100 Gbit per second \cite{Bruynsteen2022}. Both, however, have a quite complex optical subsystem. Phase noise measurements use stabilized asymmetric interferometers. Vacuum fluctuations QRNGs perform homodyne measurements implemented by a well-balanced system of a beam splitter and two photodiodes to suppress the noise of the local oscillator.

Security proofs of QRNGs based on homodyne measurements of vacuum fluctuations using a device-dependent security model are well developed. The random number output stream has been proven secure against quantum side-channels and many device imperfections have been taken into account, for instance correlations between subsequent measurements due to finite detection bandwidth~\cite{Gehring2021Homodyne-basedSide-information}. However, the potential non-Gaussian statistics of (for example) the electronic noise of the detection system and the residual relative-intensity noise of the local oscillator have not been accounted for in previous proofs which have adopted a Gaussian assumption. 
Partially this has been solved by source independent QRNGs which do not make any assumptions on the measured quantum states, however they require even more complex optical subsystems comprising two homodyne detectors for phase-diverse measurements~\cite{Marangon2017,Avesani2018} or including tap-offs for additional energy measurements~\cite{Drahi2020CertifiedLight}.

Here, we propose and experimentally demonstrate a QRNG based on the detection of vacuum fluctuations using the simplest possible photonic subsystem that consists solely of a laser and a photodiode with non-unit quantum efficiency. The quantum efficiency, strictly smaller than $100\,\%$, is the key component enabling this simple design as it effectively acts as a beam splitter through which a pure vacuum state enters the measured optical mode. This vacuum state is trusted and independent of the state emitted by the laser and is therefore our source of entropy. While the photodetector is assumed to perform a quadrature measurement on a single mode, our security analysis does not make any assumption on the emitted quantum state in the frequency sidebands of the laser carrier. In particular, it does not require the laser noise nor the electronic noise of the detector to be Gaussian.

We perform an experimental demonstration of this simple optical scheme based on cost-efficient off-the-shelf components: A vertical-cavity surface-emitting laser (VCSEL) and a silicon pin photodiode. VCSEL technology has been developed in the past decades and VCSEL diodes are becoming a low-cost alternative to traditional single-frequency lasers such as distributed feedback and distributed Bragg reflection lasers. There are very few reported cases of VCSEL usage for randomness generation \cite{Shakhovoy2021, Quirce2022}.

In our experiment, we characterized
the parameters in our stochastic model rigorously and thereby achieve composable security. Real-time randomness extraction based on Toeplitz hashing, implemented on a field-programmable gate array (FPGA), completes 
our experimental demonstration. The achieved output rate of random numbers is 3.41 Gbit/s.

The simplicity of the photonic circuit combined with the achieved high generation rate in the Gbit per second range, shows that photonic circuit integration of the optical subsystem~\cite{Abellan2016,Huang2019,Bai2021,Bruynsteen2022} is not a necessity for cost-effective optical QRNGs. At the same time, our security analysis advances the security of device-dependent QRNGs, as it allows for arbitrary noise distributions while still including quantum side-information and finite detection bandwidth.

\section{Results}

\subsection{General idea}
\begin{figure}
    \includegraphics[width=\linewidth]{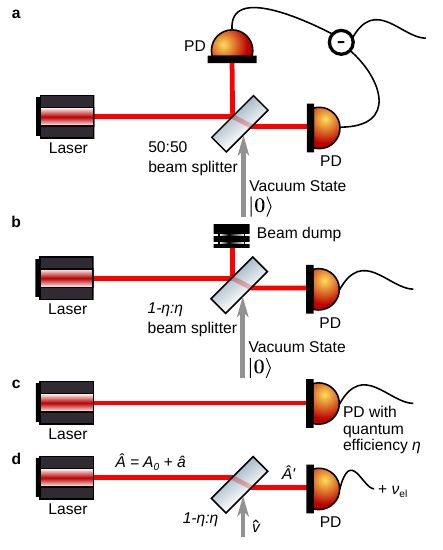} \\
    \caption{\textbf{Idea and physical model.} a) Traditional scheme for QRNGs based on the measurement of vacuum fluctuations. The vacuum state $|0\rangle$ is measured by balanced homodyne detection driven by a local oscillator from a laser. PD: photodiode. b) Our vacuum fluctuations QRNG with simpler optical schematic. The noisy laser beam first interferes with vacuum at a beam splitter (that models the quantum efficiency of the photodiode) with transmissivity $\eta$ before being detected by a single photodiode. The reflected beam off the beam splitter is safely dumped and not accessible by anyone. c) Practical implementation of the QRNG scheme depicted in b). d) Physical model of the QRNG. A mode of the laser is described by the quantum state $\hat{A}$ which is decomposed into the expectation value $A_0$ and the fluctuations $\hat{a}$. The vaccum mode is labeled $\hat{v}$ and the mode after interference is labeled $\hat{A}'$. The photodetection process adds noise $\nu_\text{el}$.}
   \label{qrng_gen_idea}
\end{figure}

Traditionally, QRNGs based on quadrature measurements of the vacuum field use balanced homodyne detection~\cite{Weedbrook2012}. As depicted in Fig.~\ref{qrng_gen_idea}a, balanced homodyne detection is performed by mixing a vacuum state with a local oscillator (LO) on a balanced beam splitter and detecting the two outputs by photodiodes whose photo currents are subtracted. The electrical output of the subtraction circuit is proportional to a quadrature of the vacuum field (if the LO power is sufficiently large) and ideally relative-intensity noise from the local oscillator is completely cancelled by the symmetry of the circuit. In practice, however, the common-mode rejection of the circuit is finite due to component imperfections and maintaining the balancing is a challenge.

Our approach simplifies the detection circuit by replacing one of the photodiodes by a beam block and by allowing for an arbitrary transmissivity $\eta < 1$ of the beam splitter. The resulting circuit is shown in Fig.~\ref{qrng_gen_idea}b. The laser output is still mixed with a vacuum state and directly detected by a photodiode. The measurement corresponds to an amplitude quadrature measurement (under the same condition of sufficiently large LO power). The mixture between laser noise and vacuum noise can be tuned by the beam splitter transmissivity.

A compact implementation of the above can be achieved by considering the quantum efficiency of the photodiode which can be modeled as a virtual beam splitter with transmissivity equal to the quantum efficiency. This is shown in Fig.~\ref{qrng_gen_idea}c. As we will see later, the entropy of the measurement results stemming from the vacuum state can be described by just two parameters: the beam splitter transmissivity and the photo diode gain (potentially containing a correction factor for finite detection bandwidth). The probability distribution of the laser noise however does not play a role and the laser noise power is only of concern as to avoid saturation of the analog-to-digital converter used to digitize the detector's output.

\subsection{Security analysis}

We present a theoretical analysis of the QRNG to bound the number of random bits that can be extracted from the measurement.
We start with declaring the assumptions underpinning the analysis.
Then we introduce the physical model and compute the min-entropy of the measurement, conditioned on quantum side information.
Finally, we use a stochastic model to link the experimental parameters to the min-entropy.

Our analysis is based on the following assumptions:
\begin{itemize}
    \item[A0] Quantum mechanics is a reliable theory.
    \item[A1] The quantum system that is measured by the photodiode is in a stationary state for the time it requires to collect enough measurement results to perform randomness extraction. Before the measurement the quantum system first experiences a pure and trusted loss as a consequence of the non-unity efficiency of the photodiode.
    \item[A2] The experimental setup implements a quadrature measurement on a single longitudinal mode. The measurement outcome is linear in the quadrature and the map from quadrature to output voltage is constant over the aforementioned time scales.
\end{itemize}

The state measured at a given time can be completely general, as long as it is a 
single mode.
In particular, we do not require that the laser or detector noise is Gaussian, thus relaxing one of the assumptions used in previous works~\cite{Haw2015, Gehring2021Homodyne-basedSide-information, Bruynsteen2022}.

Next, we describe the physical model of the QRNG shown in Fig.~\ref{qrng_gen_idea}d.
The laser output can be written as a field $\hat A = A_0 + \hat a$, where $A_0$ is a complex number describing the optical carrier's classical amplitude and phase, and $\hat a$ is a bosonic (annihilation) operator that represents quantum noise from the laser or the external environment.
The field is mixed with the vacuum mode $\hat{v}$ at the virtual beam splitter placed in front of the photo diode, which models the quantum efficiency $\eta$. The output mode of the beam splitter towards the photo diode is
\begin{align}
    \hat A' & = \sqrt{\eta} A_0 + \sqrt{\eta} \, \hat a + \sqrt{1-\eta} \, \hat v = \sqrt{\eta} A_0 + \hat b \, ,
\end{align}
where $ \hat b = \sqrt{\eta} \, \hat a + \sqrt{1-\eta} \, \hat v$ and where only the second part (from the beam splitter vacuum fluctuations of $\hat v$) contributes to the extraction of random numbers. A photodiode measures the operator
\begin{align}
|\hat A'|^2   
& \simeq \eta |A_0|^2 + \sqrt{2\eta} |A_0| \hat x_\phi \, ,
\label{eq_B}
\end{align} 
where $\hat x_\phi = (e^{i\phi} \hat b^\dag + e^{-i\phi} \hat b )/ \sqrt{2}$ is the quadrature of the field that is actually measured. We made the assumption that the power $\eta|A_0|^2$ in the optical carrier is much stronger than the noise power in the sidebands which is given by the expectation value of $\hat{b}^\dagger \hat{b}$. This assumption ensures that the detector performs a quadrature measurement. The first term associated with the laser power is removed by the highpass filter at the detector output. 

The photo detection circuit adds classical noise $\nu_\text{el}$ to the measurement data (yielding another term in Eq.~(\ref{eq_B})). 
In the worst case this contribution is fully quantum correlated to the environment. Therefore we absorb it into the operator $\hat{a}$ thereby considering the electronic noise as extra laser noise.
Thus, the operator $\hat{a}$ expresses all fluctuations coming from the environment through the laser and other sources of noise.
Furthermore, the photodiode is trusted and shielded from the environment, in such a way that no information may leak through the other output port of the virtual beam splitter in Fig.~\ref{qrng_gen_idea}d.

We now deduce the extractable randomness and explore its behavior as a function of some measurable parameters. The results of our derivation are briefly presented while details can be found in the Methods. Aggregating $l$ measurements, the extractable randomness is given by the left-over hash lemma,
\begin{equation}
    \ell \ge lH_\text{min}(Y|E) - \log_2\frac{1}{2\epsilon^2_\text{hash}}\ ,
\end{equation}
where $H_\text{min}$ is the min-entropy, $Y$ is the measurement output, $E$ represents the environment, and $\epsilon_\text{hash}$ is the statistical distance between perfect random numbers and the string generated by the randomness extractor.

Let $|\Psi\rangle_{Ea}$ be the pure quantum state that describes the quantum correlations between the light emitted by the laser (represented by the field operator $\hat{a}$) and the environment. Without loss of generality we can write this state as
\begin{align}\label{state_Ea}
    |\Psi\rangle_{Ea} = \int dx \, \alpha(x) |\psi (x)\rangle_E \otimes |x\rangle_a \, ,
\end{align}
where $|x\rangle_a$ is the (generalized) eigenvector of the quadrature operator associated to mode $\hat{a}$, and $\{ |\psi (x)\rangle_E \}_{x \in \mathbb{R}}$ is a continuous family of states for the environment (not necessarily orthogonal). $\alpha (x)$ is a complex function describing the correlations and we note that $\int dx |\alpha(x)|^2 = 1$ to ensure normalization.
As discussed above, the non-unit efficiency of homodyne detection is modeled as arising from mixing with a trusted vacuum mode $v$, at a beam splitter of transmissivity $\eta$. We express the vacuum state as
\begin{align}\label{state_vacuum}
    |\phi\rangle_v = \int du \sqrt{F(u)} \, |u\rangle_v \, ,
\end{align}
where $|u\rangle_v$ is the (generalized) eigenvector of the quadrature operator for mode $v$, and $F(u) = (2\pi)^{-1/2} e^{-u^2/2}$ (variance of vacuum is normalized to unity). 

After mixing the vacuum with the laser beam, followed by detection, the classical-quantum state $\rho_{EY}$ is prepared, where $Y$ is the classical variable that describes the measurement output.
To obtain a lower bound on the min-entropy, we first extend the state $\rho_{EY}$ into $\rho_{EXY}$, where $X$ is an auxiliary (and un-observed) classical variable that carries the values of $x$ associated with the laser noise (Eq.~(\ref{state_Ea})).
Since conditioning does not increase entropy, we have
\begin{align}\label{tool-1}
H_\mathrm{min}(Y|E) \geq H_\mathrm{min}(Y|EX) \, . \end{align}
We then bound $H_\mathrm{min}(Y|EX)$ as \cite{Tomamichel2012}
\begin{align} \label{tool-2}
H_\mathrm{min}(Y|EX) \geq - \log{
\left\| 
\gamma_{EX}^{-1/2} \, \rho_{EXY} \, \gamma_{EX}^{-1/2}  \right\|_\infty } \, ,
\end{align}
where $\| \, \cdot \, \|_\infty$ is the operator norm (i.e., the largest eigenvalue).
A lower bound is obtained for any choice of the state $\gamma_{EX}$ in Eq.~(\ref{tool-2}). In particular, we obtain our bound from the choice
\begin{align}\label{qw8nc}
\gamma_{EX} = \int dx |\alpha(x)|^2  
 |\psi (x)\rangle_E \langle \psi(x)| \otimes |x\rangle_X \langle x| \, .
\end{align}

For ideal homodyne detection with infinite range and precision, and by using the expression for $\gamma_{EX}$ in Eq.~(\ref{qw8nc}), the explicit evaluation of Eq.~(\ref{tool-2}) yields
\begin{align}
H_\mathrm{min}(Y|E) \geq 
- \log\left( \frac{1}{ 2\pi  g \sqrt{1-\eta}} \right) \, ,
\label{eq_min_entr_basic}
\end{align}
where $g$ is the gain of the measurement device, i.e.\ the standard deviation of the vacuum noise.
%
This bound is universal as it does not depend on $\Psi$: it only depends on the features of the trusted measurement device, and vacuum mode used to model non-unit efficiency.
When taking into account that the output measurement is discretized using an analog-to-digital converter (ADC), that the ADC has finite range, and it is subject to digitization errors, we obtain
\begin{align}
 \nonumber   H_\mathrm{min}(Y|E) 
    & \geq - \log 
    \mathrm{erf}\left( \frac{R}{g d \sqrt{2 (1-\eta)}} \right) 
\\ & 
\phantom{=}~
+ \log{P} - b_\mathrm{ADC} \, ,
\label{eq_min_entr1}
\end{align}
where erf is the error function,
$R$ is the range of the ADC, 
$d$ is the number of bins, 
and $P$ is the probability that a measurement output falls inside the ADC range $(- R, R)$. 
Note that if $g$ is measured in units of ADC bins, then usually $R/d = \tfrac{1}{2}$. 
Boundary bins are discarded in our model of the ADC, which reduces the min-entropy by $\log{P}$ bits. 
Finally, the term $b_\mathrm{ADC}$ comes from the digitization errors that are observed in the ADC.

\begin{figure}
    \centering
    \includegraphics{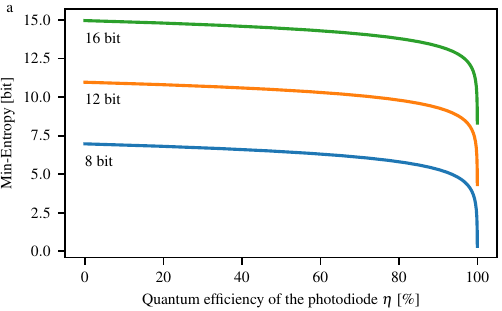} \\
    \includegraphics{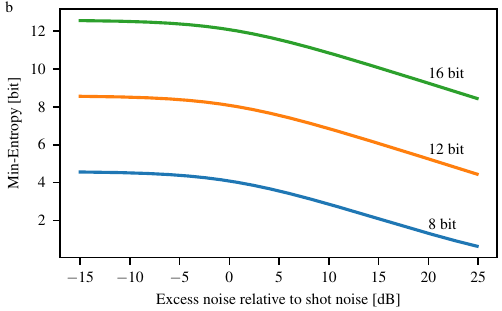} \\
    \caption{\textbf{Theoretical investigations.} a) Min-entropy bound in Eq.~(\ref{eq_min_entr1}) versus the quantum efficiency of the photodiode for an ideal 8, 12 and 16 bit ADC with excess noise following a Gaussian distribution and a detected noise power 5 dB larger than the vacuum noise. The detection gain has been optimized without practical constraints to reach the absolute maximum of the min-entropy. b) Min-entropy versus detected Gaussian excess noise for a photodiode with $80\,\%$ quantum efficiency. The value of the detection gain $g$ has been suitably chosen in order to not saturate the ADC, in such a way that 6 standard deviations of the detected signal fit into the ADC range.}
    \label{fig:theory}
\end{figure}

Equation~(\ref{eq_min_entr1}) shows that there exists an optimal value of $g$.
In fact, the term with $R/g$ increases with increasing $g$ (higher signal level on the ADC input), while the term with $P$ decreases (as more samples fall out of the range). Considering the optimal values for $g$, Fig.~\ref{fig:theory}a shows the min-entropy versus the quantum efficiency of the photodiode for three different ADC resolutions. Clearly the min-entropy vanishes for an ideal photodiode with 100\,\% efficiency, and in general is a monotonically decreasing function of the efficiency.
However, in practice a low quantum efficiency is impractical as it requires a laser with a larger output power to overcome the electronic noise of the detector. Furthermore, the optimal values for $g$ are quite large which leads to ADC saturation, a situation that is better avoided in practice.

This is taken into account in Fig.~\ref{fig:theory}b where we limit the gain such that 6 standard deviations of the signal are contained within the ADC range which corresponds to $P$ being constant. The graph shows the min-entropy versus the excess noise relative to shot noise for a photodiode efficiency of 80\,\%. As can be seen in the figure the min-entropy decreases with increasing excess noise which is due to a decrease of the gain to maintain that 6 standard deviations fit into the ADC range. Excess noise due to relative intensity noise of the laser is the most important source of noise for this type of QRNG since it is not suppressed as in regular homodyne detection. This means that the quality of the laser becomes an important figure of merit for an experimental realization, however as can be seen in the graph, a decent amount of excess noise can be tolerated.

Finally, we consider the effect of finite detection bandwidth. While the vacuum state used to model the non-unit efficiency of the detector has, in principle, infinite bandwidth, the detector's frequency response allows us to measure only a limited bandwidth.
For a detector with linear response, the signal recorded at time $t$ contains linear terms depending on past values at times $t'<t$. 
Within this linear model, the min-entropy can still be bounded and we obtain an expression formally analogous to Eq.~(\ref{eq_min_entr1}), with $g$ replaced by $g \chi_0^u$, where the parameter $\chi_0^u$ accounts for the finite bandwidth. Details are given in the Methods.

To summarize the theory of our QRNG, we found that the min-entropy can be computed using the parameters $g\chi_0^u$, $\eta$, and $P$, which in an experimental realization have to be determined with a confidence level that is quantified by $1-\varepsilon$ where $\varepsilon$ is the error parameter.
This parameter is required to be fairly small (about $10^{-10}$ or smaller) for most applications in cryptography.

\subsection{Experimental setup}

\begin{figure}[ht]
 \includegraphics{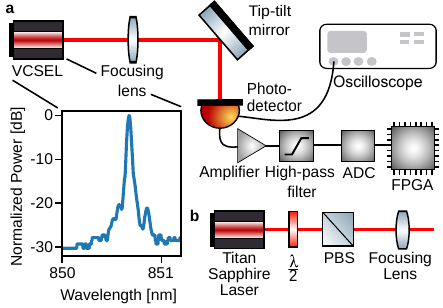} \\
 \includegraphics[width=0.9\linewidth]{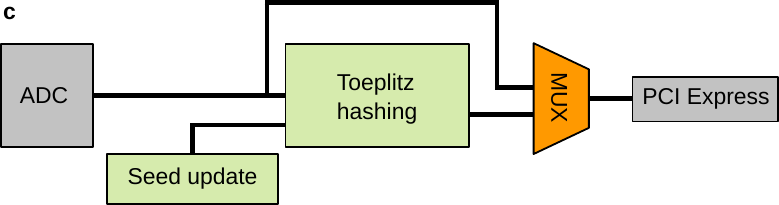}
 \caption{\textbf{Experimental setup.} a) The experimental setup consists of free-space optical components and electronics. The red lines designate optical signals, and the black lines show electrical connections between components. A laser beam from a VCSEL is focused with a lens to a single photo diode. The photo current is converted to voltage with a transimpedance amplifier, highpass filtered and sampled by an ADC read out by an FPGA. The DC output voltage of the detector was monitored on an oscillosocpe. The inset shows an optical power spectrum of the VCSEL output measured with an optical spectrometer. {b)} For the system characterization stage we used a slightly modified setup with a Titan Sapphire laser and a half-wave plate ($\lambda/2$)/polarizing beamsplitter (PBS) combination to control the optical power. c) FPGA firmware block diagram. The ADC output is fed either to the Toeplitz hashing module in case of randomness extraction, or is directly offloaded to the host computer via PCI Express. The Toeplitz matrix seed can be updated via software.}
 \label{experimental_setup}
\end{figure}

The experimental setup is shown in Fig.~\ref{experimental_setup}. We implemented the laser and the photodiode onto two separate printed circuit boards to simplify the proof-of-concept demonstration. The VCSEL emitted light at around 850 nm into a single longitudinal mode below an output power of 2 mW (see inset in Fig.~\ref{experimental_setup}). The laser beam had a Gaussian spatial profile with 7 degrees divergence angle. A focusing lens and a steering mirror were used to focus the beam onto the photo detector --- however these optical components are not strictly necessary. The photodiode was a silicon pin diode and allowed our custom-made photo detection circuit to reach a bandwidth of about 250 MHz. We separated low frequency components due to the first term in Eq.~(\ref{eq_B}) from high frequency components related to the quadrature measurement, due to the second term in Eq.~(\ref{eq_B}). The cutoff frequency of the highpass filter for the quadrature measurements was set to 5 MHz to suppress relative intensity noise during system characterization, as we describe below.  
The output was sampled with a 16-bit ADC at 1 GS/s which was read out by a field-programmable gate array (FPGA).

The functionality of the FPGA is shown in Fig.~\ref{experimental_setup}c. The ADC output was fed into a Toeplitz hashing module which implemented randomness extraction. The final random numbers were offloaded via PCI Express to the host computer. A multiplexer allowed for bypassing the hashing module during the system characterization.

\subsection{System characterization}

\begin{figure*}
    \centering
    \includegraphics{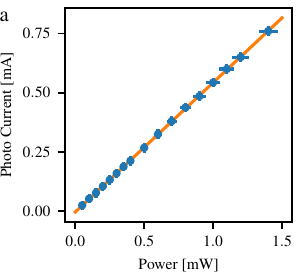} 
    \includegraphics{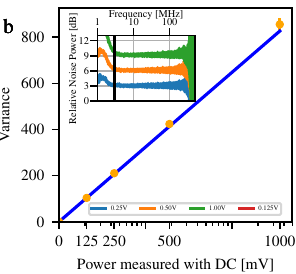}
    \includegraphics{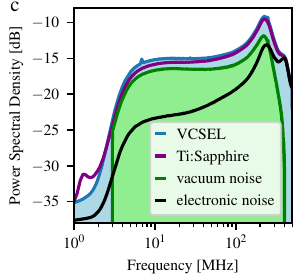}\\
    \caption{\textbf{System characterization} a) Photo diode responsivity measurement. b) Verification of shot-noise limited behaviour of the Titan Sapphire laser. Inset: Determination of the frequency from which the laser was shot-noise limited.
    c) Measurement for the determination of the finite bandwidth--gain product. The PSD of the vacuum noise (green) has been calculated by subtracting the PSD of the electronic noise of the detector (black) from the PSD of the noise measured with the Titan Sapphire laser (purple). Since Titan Sapphire laser is not shot noise-limited below around 3 MHz the PSD of the vacuum noise has been set to 0 below that frequency. For comparison we show the PSD of the noise measured with the VCSEL laser (blue).
    }
    \label{fig:characterization}
\end{figure*}

We will now discuss the characterization of the parameters of the stochastic model required to calculate the extractable min-entropy in our experiment.
The four parameters are the quantum efficiency of the photodiode $\eta$, the finite-bandwidth gain $g\chi_0^u$, the ratio of samples falling into the ADC range $P$, and the reduction $b_\text{ADC}$ from the digitization error of the ADC.

Figure~\ref{fig:characterization}a shows the results of the characterization of the photodiode's quantum efficiency. To obtain these results, we measured the photo current versus the optical power using a Ti:Sapphire laser tuned to 850 nm -- identical to the wavelength of the VCSEL. The optical power was controlled by a half-waveplate in combination with a polarizing beam splitter as depicted in Fig.~\ref{experimental_setup}b, and measured with a commercial power meter. The power meter had an error of $5\,\%$. The responsivity $K_{\lambda}$ of the photodiode in Ampere per Watt was determined by a linear fit and then converted to quantum efficiency via
\begin{align}
    \eta=\frac{e\lambda}{K_\lambda h c}\ ,
\end{align}
where $e$ is the electron charge, $\lambda$ is the wavelength, $h$ is Planck's constant and $c$ is the speed of light. We determined the quantum efficiency to $\eta = 0.796 (2)$ and bound it by $\eta < 0.81$ which corresponds to a confidence level of $(1-\varepsilon_\eta)$ with  $\varepsilon_\eta=10^{-11}$.

The next parameter to be bounded is the gain-bandwidth product, $g \chi_0^u$. The first factor in this product is the gain of the measurement apparatus while the second quantifies the bandwidth of the detector. $\chi_0^u$ takes the value 1 for a detector of infinite bandwidth. The gain is a combination of optical gain that depends on the optical power of the carrier and electronic gain of the detection circuit and the ADC.
We use the high-quality Ti:Sapphire laser to measure the gain-bandwidth product.  
This allows us to determine the power spectral density (PSD) of the measured vacuum fluctuations beyond the sideband frequency at which the laser is shot noise limited. Below that frequency, we set the PSD of the vacuum fluctuations to zero as our method does not allow us to determine the contribution of the vacuum fluctuations to the total noise power. Due to the finite bandwidth of the detector, the PSD is expected to be colored. As vacuum fluctuations have a Gaussian distribution we can compute the conditional standard deviation from the PSD $f_\text{vac}$, which in turn gives the factor $g \chi_0^u$ according to Eq.~(\ref{cond_gain_eq}) --- see Methods, Section~\ref{sec:finite_detection_bandwidth}.

To implement this procedure we first determined the frequency above which the Ti:Sapphire laser was shot-noise limited. For this purpose we plotted the frequency spectrum of the noise power for 0.4, 0.8 and 1.6 mW relative to the noise power recorded with 0.2 mW. The result is shown in the inset of Fig.~\ref{fig:characterization}b. In the ideal case of shot-noise limitation, the noise power will increase by 3 dB when doubling the input power, and this appears to be the case for frequencies larger than 3 MHz. 
This was confirmed by integrating the frequency spectra from 3 to 500 MHz and plotting the noise variance versus the optical power. This is shown in Fig.~\ref{fig:characterization}b where the blue line shows the expected behavior for a shot-noise limited laser.

Finally, Figure~\ref{fig:characterization}c shows the PSD of the vacuum noise (green trace). This was obtained by subtracting the PSD of the electronic noise of the detector (black trace) from the measured PSD (purple trace). As mentioned above we set the PSD of the vacuum noise to 0 below 3 MHz. From this we compute $g \chi_0^u$ according to Methods, Section~\ref{sec:finite_detection_bandwidth}. The result is listed in Table~\ref{table1}. For comparison we also plotted the PSD of the noise measured with the VCSEL instead of the Ti:Sapphire laser (blue trace) which shows that relative intensity noise of the VCSEL is not far from shot-noise.

Next, we bound $P$ by counting the out-of-bounds samples for $10^9$ measurements and we compute confidence intervals using the Hoeffding tail bound. Details are given in the Methods, Section~\ref{sec:out_of_bound_samples}.

At last, we determined the reduction of the min-entropy due to the digitization error of the ADC. This is presented in Methods, Section~\ref{sec:ADC digitization error} where we adopted the method presented in Ref.~\citenum{Gehring2021Homodyne-basedSide-information} for our QRNG. The measurement results are described in Methods, Section~\ref{sec:measurement_of_ADC_digitization_error}. We note however that this method is quite conservative and therefore the measured reduction is as large as 7.8 bits. In Ref.~\citenum{Bruynsteen2022} another method is presented which leads to more realistic reductions. It is however not applicable to our ADC due to its interleaving nature. 

All parameters of the characterization are summarized in Table~\ref{table1}, yielding a final min-entropy of 3.67 bits per measurement. Using real-time randomness extraction implemented on the FPGA, we reached a speed of 3.41 Gbit/s.

\begin{table*}[!t]
\begin{tabular}{l|p{6cm}}
    \textbf{Parameter} & \textbf{Value}\\
    \hline
    Optical power $P_\mathrm{in}$ detected by photodiode & 662 $\mu$W  \\
    Standard deviation of VCSEL noise, $\sigma_\mathrm{VCSEL}$ & $5043$ [ADC units]. About 6.5 standard deviations fit into the ADC range.\\
    \hline
    Quantum efficiency $\eta$ & $< 0.81$ \\ 
    \hline
    Conditional standard deviation of the vacuum noise $g\chi_0^u \pm\Delta (g\chi_0^u )$ & $2581 \pm (-181) (+194)$\\
    \hline
    Ratio samples falling into ADC range, $P\pm \Delta P$ &  $1 - 10^{-4}$ \\
    Total number of counted samples, $N$ & $1 \times 10^{9}$\\
    Number of out-of-interval samples, $N-N_1$ & $0$ \\
    \hline
    Min-entropy (perfect ADC), $H'_\mathrm{min}$ & 
    $11.47$ bits per 16-bit sample\\
    Min-entropy reduction due to ADC imperfections & $7.80$ bits per 16-bit sample \\
    Final min-entropy, $H_\mathrm{min}$ & $3.67$ bits per 16-bit sample \\
    \hline
    Randomness extractor Toeplitz matrix dimensions, $k \times l$ & $1680 \times 7872$ bits  \\
    Random number output rate & 3.41 Gb/s \\
    \hline
    Failure probability for parameter estimation, $\varepsilon_\mathrm{param}$ & $10^{-10}$\\
    Failure probability for ADC digitization estimation, $\varepsilon_\mathrm{ADC}$ & $6\times 10^{-5}$\\
    Failure probability of hashing, $\varepsilon_\mathrm{hash}$ &
    $10^{-17}$
\end{tabular}
\caption{The table shows the values of the relevant parameters. Where applicable, the values are shown with their corresponding confidence intervals, with the confidence levels not greater than $\varepsilon_\mathrm{param}$.}
\label{table1}
\end{table*}

\section{Discussion}
 In conclusion, we demonstrated a simplified optical scheme for QRNGs based on quadrature measurements. We combine the simplicity of a direct detection optical setup with a rigorous security analysis where the exploited quantum state is a pure state. The security analysis relies on several assumptions. Compared to  previous works on quadrature detection of vacuum states~\cite{Haw2015,Gehring2021Homodyne-basedSide-information}, we abandon the assumption of Gaussianity. This is an important step as any assumption on the distribution of a certain noise process is very difficult to verify for any practical implementation.

Our theoretical method should also be applicable to regular balanced homodyne detection of vacuum states which suppresses the relative intensity noise of lasers and is therefore expected to deliver higher min-entropy. Reducing the number of assumptions in device-dependent QRNGs is an important task towards making QRNGs more practical and eventually to achieve certification.

 We experimentally demonstrated our scheme which is the first example of low-cost VCSEL technology being employed for multi-Gbps real-time true randomness generation. It furthermore shows a path towards compact QRNGs that do not rely on photonic integrated circuits.

Future work could consider removing the single-mode assumption as many VCSEL do not emit into a single frequency mode, but have to be designed specifically to do so. Removing this assumption would allow for a wider range of cost-effective lasers driving the entropy source. This applies not only to VCSEL but also other lasers.

\section{Methods}

\subsection{Detailed derivation of the min-entropy}

The state $|\Psi\rangle_{Ea}$ describes the (quantum) correlations between the measured longitudinal mode $a$ and the environment $E$.
The representation in
Eq.~(\ref{state_Ea}) is completely general and includes all possible classical and quantum correlations with the environment. In fact, classical correlations can be represented as quantum correlations via purification.
A quadrature measurement on mode $a$ has non-unit efficiency $\eta \in (0,1)$, which is modeled as mixing with the vacuum mode at a beam splitter of transmissivity $\eta$.
The latter is formally represented as a unitary operator $U_{av}$ and transforms the state as follows
\begin{align}\label{o8WNCD3}
    |\Psi\rangle_{Ea} \otimes |\phi\rangle_v \to \left( I_E \otimes U_{av} \right) |\Psi\rangle_{Ea} \otimes |\phi\rangle_v  \, ,
\end{align}
where $|\phi\rangle_v$ is the vacuum state [Eq.~(\ref{state_vacuum})], $I_E$ is the identity map on the environment, and the unitary $U_{av}$ acts on systems $a$ and $v$.
To write this transformed state explicitly, we first recall the action of a beam splitter unitary on the generalized eigenstates of the quadrature operators,
\begin{align}
   \nonumber & U_{av} |x\rangle_a \otimes |u\rangle_v \\
 & = |\sqrt{\eta} \, x + \sqrt{1-\eta} \, u \rangle_b \otimes | \sqrt{\eta} \, u - \sqrt{1-\eta} \, x \rangle_v \, .
\end{align}

Applying this general rule to Eq.~(\ref{o8WNCD3}) and using the expansions in
Eqs.~(\ref{state_Ea}) and (\ref{state_vacuum}) we obtain the state
\begin{align}
 \nonumber & \int dx \, du \, \alpha(x) \sqrt{F(u)} \, |\psi (x)\rangle_E \otimes U_{av}  |x\rangle_a \otimes  |u\rangle_v \\ 
\nonumber    & = \int dx \, du \, \alpha(x) \sqrt{F(u)} \, |\psi (x)\rangle_E   \\ 
& \phantom{==}~
    \otimes |\sqrt{\eta} \, x + \sqrt{1-\eta} \, u\rangle_b 
    \otimes  |\sqrt{\eta} \, u - \sqrt{1-\eta} \, x \rangle_v
\, .
\end{align}

First, we note that the mode $v$ at the output of the beam splitter is not measured and is shielded from the environment, therefore we can trace over it.
The result is a mixed state
\begin{align}
\rho_{Eb} & = \int dx \, dx' \, du \, du' \, \alpha(x)\alpha^*(x') \sqrt{F(u)F(u')} \nonumber \\ 
& \phantom{==}~
    \times
    \delta(\sqrt{\eta} \, u - \sqrt{1-\eta} \, x - \sqrt{\eta} \, u' + \sqrt{1-\eta} \, x') \nonumber \\ 
& \phantom{==}~
    \otimes |\psi (x)\rangle_E\langle \psi (x')| \nonumber \\ 
& \phantom{==}~
    \otimes |\sqrt{\eta} \, x + \sqrt{1-\eta} \, u\rangle_b \langle \sqrt{\eta} \, x' + \sqrt{1-\eta} \, u'| 
\, .
\end{align}

Second, the quadrature of the mode $b$ is measured, yielding as output the random variable $Y$.
The values assumed by $Y$ are proportional to $\sqrt{\eta} \, x + \sqrt{1-\eta} \, u$ with proportionality factor given by the gain:
\begin{align}
y =  g \left( \sqrt{\eta} \, x + \sqrt{1-\eta}\,  u \right) \, .
\label{HPic}
\end{align}
The state after the measurement is the classical-quantum state $\rho_{EY}$, where $Y$ is the classical part, and $E$ carries all quantum as well as classical correlations with the environment.
We have
\begin{align}
\rho_{EY} & = \int dx \, dx' \, du \, du' \, dy \, 
\alpha(x)\alpha^*(x') \sqrt{F(u)F(u')} \nonumber \\ 
& \phantom{==}~
    \times
    \delta(\sqrt{\eta} \, u - \sqrt{1-\eta} \, x - \sqrt{\eta} \, u' + \sqrt{1-\eta} \, x') \nonumber \\
&\phantom{==}~
    \times
    \delta(y- g\sqrt{\eta} \, x -g \sqrt{1-\eta} \, u) \nonumber \\
&\phantom{==}~
    \times
    \delta(y- g\sqrt{\eta} \, x' -g \sqrt{1-\eta} \, u')
    \nonumber \\
& \phantom{==}~
    \otimes |\psi (x)\rangle_E\langle \psi (x')| 
    \otimes |y \rangle_Y
    \langle y |
    \, .
\end{align}

Note that since the delta functions enforce the constraints $\sqrt{\eta} \, x - \sqrt{1-\eta} \, u = \sqrt{\eta} \, x' - \sqrt{1-\eta} \, u'$ and 
$\sqrt{\eta} \, x + \sqrt{1-\eta} \, u = \sqrt{\eta} \, x' + \sqrt{1-\eta} \, u'$, we also have that $x=x'$ and $u=u'$. Therefore the above state reads 
\begin{align}
\rho_{EY} & = \int dx \, du \, dy \, |\alpha(x)|^2 F(u) 
    \, \delta(y- g\sqrt{\eta} \, x -g \sqrt{1-\eta} \, u) \nonumber \\
& \phantom{==}~
    \times |\psi (x)\rangle_E\langle \psi (x)| 
    \otimes |y \rangle_Y
    \langle y |
    \, .
    \label{state1}
\end{align}

Our goal is to bound the conditional min-entropy $H_\mathrm{min}(Y|E)_\rho$. 
To obtain a computable lower bound we extend the state into 
\begin{align}
 \nonumber &   \rho_{EXY}
 \\ \nonumber & = \int dx \, du \, dy \, |\alpha(x)|^2 F(u)
    \delta (y - g\sqrt{\eta} \, x - g\sqrt{1-\eta} \, u)
     \\  
    &
    \phantom{==}~\times
    |\psi (x)\rangle_E \langle \psi(x)| 
    \otimes |x\rangle_X \langle x| 
    \otimes |y\rangle_Y\langle y | \, ,
\end{align}
where we have introduced the classical variable $X$, which carries the values of $x$.
As conditioning decreases entropy, the following lower bound holds
\begin{align}
    H_\mathrm{min}(Y|E)_\rho \geq H_\mathrm{min}(Y|EX)_\rho \, .
\end{align}

From the definition of min-entropy we obtain a lower bound for any density matrix $\gamma_{EX}$ \cite{Tomamichel2012},
\begin{align}
H_\mathrm{min}(Y|EX)_\rho \geq - \log{
\left\| 
\gamma_{EX}^{-1/2} \, \rho_{EXY} \, \gamma_{EX}^{-1/2}  \right\|_\infty } \, ,
\end{align}
%
where
\begin{align}
& \| \gamma_{EX}^{-1/2} \, \rho_{EXY} \, \gamma_{EX}^{-1/2} \|_\infty \nonumber \\
& = \bigg\| \int dx \, du \, dy \,
|\alpha(x)|^2 F(u) \, 
\delta (y - g\sqrt{\eta} \, x - g\sqrt{1-\eta} \, u ) 
 \nonumber\\ 
&
\phantom{===}~ \times 
\gamma_{EX}^{-1/2} 
\left( 
|\psi (x)\rangle_E \langle \psi(x)| 
\otimes |x\rangle_X \langle x| \right) \gamma_{EX}^{-1/2} 
\nonumber \\ & 
\phantom{===}~
\otimes |y\rangle_Y\langle y | \bigg\|_\infty \, .
\end{align}

By putting
\begin{align}
\gamma_{EX} = \int dx |\alpha(x)|^2  
 |\psi (x)\rangle_E \langle \psi(x)| \otimes |x\rangle_X \langle x| \, , 
\end{align}
we obtain
\begin{align}
\nonumber & \left\| \gamma_{EX}^{-1/2} \, \rho_{EXY} \, \gamma_{EX}^{-1/2} \right\|_\infty
\\ \nonumber
& = 
\bigg\| \int dx \, du \, dy \,
F(u) \, \delta (y - g\sqrt{\eta} \, x - g\sqrt{1-\eta} \, u )
\\
&
\phantom{===}~ 
\times  
|\psi (x)\rangle_E \langle \psi(x)| \otimes |x\rangle_X \langle x| 
\otimes  |y \rangle_Y\langle y | \bigg\|_\infty \\
& = 
\sup_{x,y} \int du \, F(u) \, \delta (y - g\sqrt{\eta} \, x -
g\sqrt{1-\eta} \, u ) \, .
\end{align}
With a change of variables this becomes
\begin{align}
\nonumber & \left\| \gamma_{EX}^{-1/2} \, \rho_{EXY} \, \gamma_{EX}^{-1/2} \right\|_\infty \\
& = \frac{1}{g\sqrt{1-\eta}} \,
\sup_{x,y} \int du \, F(u) \, \delta\left( 
u - \frac{y-g \sqrt{\eta}\, x }{ g \sqrt{1-\eta}}
\right) \\
& = \frac{1}{g\sqrt{1-\eta}} \,
\sup_{x,y} 
F\left( \frac{y - g \sqrt{\eta}\, x }{ g \sqrt{1-\eta}} \right) \\
& = \frac{1}{g\sqrt{1-\eta}} \,
\sup_{u} 
F(u) \, .
\label{final-simple}
\end{align}

In conclusion, we have lower-bounded the min-entropy of the measurement output $Y$ in terms of that of the trusted vacuum mode $v$.
For pure vacuum, $F(u) = (2\pi)^{-1/2} e^{-u^2/2}$, and we obtain
\begin{align}\label{c9m43}
H_\mathrm{min}(Y|E)_\rho \geq 
- \log\left( \frac{1}{ 2\pi  g \sqrt{1-\eta}} \right) \, .
\end{align}

\subsubsection{ADC - intermediate model}

The analysis gets more involved when taking into account that the (in principle continuous) measurement output is in fact discretized using an ADC.
As a first step in our analysis, consider a simplified model of ADC where the continuous output $y$ is discretized according to the mapping $y \to j$ for any $y \in I_j$, where 
\begin{align}
I_j = [-R + 2Rj/d, -R + 2R(j+1)/d)
\end{align}
for $j=-\infty,\dots, \infty$.
This model represents an ADC with infinite range and bin size $2R/d$.
From Eq.~(\ref{state1}) we obtain the probability of measuring $j$,
\begin{align}
\nonumber  &  p_{\bar Y}(j)
 \\    & = \int_{I_j} dy 
    \int dx \, du \,  
    |\alpha(x)|^2 F(u) \,
    \delta ( y - g\sqrt{\eta} \, x - g\sqrt{1-\eta} \, u ) 
     \, ,
\end{align}
where we have introduced the discrete random variable $\bar Y$, which carries the value of $j$.
The conditional state of the environment, given $j$, is
\begin{align}\label{state1-0}
\nonumber &    \rho_{E}(j) \\ \nonumber
& = \frac{1}{p_{\bar Y}(j)} \, \int_{I_j} dy 
    \int dx \, du \, 
    \delta ( y - g\sqrt{\eta} \, x - g\sqrt{1-\eta} \, u ) \\ 
    & \phantom{======}~
    \times |\alpha(x)|^2 F(u) 
    |\psi (x)\rangle_E \langle \psi(x)|  \, .
\end{align}

We can then write the joint classical-quantum state
\begin{align}\label{state2}
 \rho_{E\bar Y}  & = \sum_{j=-\infty}^\infty p_{\bar Y}(j) \rho_{E}(j) \otimes |j\rangle_{\bar Y}\langle j | \\ 
    & = \sum_{j=-\infty}^\infty 
    \int_{I_j} dy 
    \int dx \, du \,  
    \delta (y - g\sqrt{\eta} \, x - g\sqrt{1-\eta} \,u ) \nonumber \\  
        & \phantom{======}~
    \times |\alpha(x)|^2 F(u) 
    |\psi (x)\rangle_E \langle \psi(x)| 
    \otimes |j\rangle_{\bar Y}\langle j | \, .
\end{align}

Proceeding as in the previous section, we obtain
\begin{align}
\nonumber & \left\| \gamma_{EX}^{-1/2} \, \rho_{EX \bar Y} \, \gamma_{EX}^{-1/2} \right\|_\infty 
\\ \nonumber 
& = 
\bigg\| \sum_j \int_{I_j} dy \int dx \, du \, 
\delta (y - g\sqrt{\eta} \, x - g\sqrt{1-\eta} \, u )
\\ \nonumber 
& \phantom{======}~
\times 
 F(u) 
    |\psi (x)\rangle_E \langle \psi(x)| \otimes |x\rangle_X \langle x | 
    \otimes |j\rangle_{\bar Y}\langle j | \bigg\|_\infty \\ \nonumber
& = 
\sup_{j,x} 
\int_{I_j} dy 
\int du \,  
\delta (y - g\sqrt{\eta} \, x - g\sqrt{1-\eta} \, u )
 F(u) \\ \nonumber
& = 
\frac{1}{g\sqrt{1-\eta}}
\sup_{j,x} 
\int_{I_j} dy 
\int du \,
\delta\left(u-\frac{y - g\sqrt{\eta} \,  x}{g\sqrt{1-\eta}}\right) \\ \nonumber
& \phantom{==========}~ 
\times F\left( \frac{y - g\sqrt{\eta} \, x}{g\sqrt{1-\eta}} \right) \\ 
 & = 
\frac{1}{g\sqrt{1-\eta}} \, 
\sup_{j,x} 
\int_{I_j} dy \,  
 F\left( \frac{y - g\sqrt{\eta} \, x}{g\sqrt{1-\eta}} \right) \, .
\end{align}

For pure vacuum, $F(u) = (2\pi)^{-1/2} e^{-u^2/2}$, and we obtain
\begin{align}
\nonumber & \left\| \gamma_{EX}^{-1/2} \, \rho_{EX \bar Y} \, \gamma_{EX}^{-1/2} \right\|_\infty 
\\  \nonumber
& =
 \frac{1}{g\sqrt{1-\eta}} \, 
\int_{-R/d}^{R/d} dy \,  
 F\left( \frac{y}{g\sqrt{1-\eta}} \right) \\
 & =
\mathrm{erf}\left( \frac{R}{g d \sqrt{2 (1-\eta)}} \right) \, .
\end{align}
This finally yields the desired min-entropy bound:
\begin{align}\label{min-E-2}
H_\mathrm{min}(Y|E)_\rho \geq - \log 
\mathrm{erf}\left( \frac{R}{g d \sqrt{2 (1-\eta)}} \right) 
\, .
\end{align}

\subsubsection{ADC - full model}

In practice, the ADC has only a finite range $R$. Typically, values that fall outside of the range are collected in a pair of boundary bins.
To simplify the theoretical analysis, here we follow a different approach and discard all the data points that would go in the boundary bins.
The state of interest can then be obtained from $\rho_{E\bar Y}$ in Eq.~(\ref{state2}) by applying the projector operator $\Pi = \sum_{j=0}^{d-1} |j\rangle_{\bar Y}\langle j |$, which maps into the space of allowed values of $j$,
\begin{align}\label{state3}
  \sigma_{{E\bar Y}} & = \frac{1}{P} \Pi \rho_{E\bar Y} \Pi \\ \nonumber
    & = \frac{1}{P} \sum_{j=0}^{d-1} 
    \int_{I_j} dy 
    \int dx \, du \,
    \delta ( y - g\sqrt{\eta} \, x - g\sqrt{1-\eta} \, u ) \\ 
    & \phantom{======}~
    \times |\alpha(x)|^2 F(u) 
    |\psi (x)\rangle_E \langle \psi(x)| 
    \otimes |j\rangle_{\bar Y}\langle j | \, ,
\end{align}
where the normalization coefficient $P$ is the probability of obtaining a measurement result within the range $[-R,R]$.

Lemma $1$ of Ref.~\cite{Lupo2018} allows us to bound the min-entropy of $\sigma_{E\bar Y}$ in terms of that of $\rho_{E\bar Y}$ in Eq.~(\ref{min-E-2}) by paying a penalty equal to $\log{P}$:
\begin{align}
 \nonumber  H_\mathrm{min}(\bar Y|E)_\sigma 
     & \geq H_\mathrm{min}(\bar Y|E)_\rho + \log{P} \\ 
    & \geq - \log 
\mathrm{erf}\left( \frac{R}{g d \sqrt{2 (1-\eta)}} \right) 
+ \log{P} \, .
\end{align}
Note that the factor $P$ needs to be estimated from the experimental data. 
For example, if the state of mode $b$ is Gaussian with variance $w$ then $P = \mathrm{erf}\left( \frac{R}{g\sqrt{2 w}} \right)$ and
\begin{align}
 \nonumber &   H_\mathrm{min}(\bar Y|E)_\sigma 
     \geq 
    \\ & - \log 
\mathrm{erf}\left( \frac{R}{g d \sqrt{2 (1-\eta)}} \right) 
+ 
\log 
\mathrm{erf}\left( \frac{R}{g\sqrt{2 w}} \right) 
\, .
\end{align}

\subsection{ADC digitization error}
\label{sec:ADC digitization error}

ADCs are not ideal devices, and are subject to digitization error.
Following Ref.~\cite{Gehring2021Homodyne-basedSide-information} we model the digitization error by introducing
a classical noise variable $C$, with associated probability distribution $p_C$.
The actual ADC output is thus obtained from the ideal, noiseless output $j$ as $f = f(j,c)$, for some (unknown, possibly non-linear) function $f$.

Within this model, the quantum side information about the output $F$ of the noisy ADC is described by the classical-quantum state
\begin{align}
\nonumber & \rho_{E F C}  \\ & = \sum_{jc} p_{\bar Y}(j) p_C(c)
\rho_E(j) 
\otimes |f(j,c)\rangle_F \langle f(j,c)|
\otimes  |c\rangle_C \langle c| \, ,
\end{align}
where we have introduced a dummy quantum register $C$ to keep track of the classical noise.
Of particular interest is the set $J_f$, defined as the set of values of $j$ such that $f(j,c)=f$ for some $c$.
Including the ADC noise as a side channel yields an additive correction to the min-entropy~\cite{Gehring2021Homodyne-basedSide-information},
\begin{align}
H_\mathrm{min} ( \bar Y | E C ) 
& \geq 
H_\mathrm{min} ( \bar Y | E ) 
- \log{ \sup_{f}
|J_{f}|  } \, .
\end{align}

In conclusion, when compared with an ideal noiseless ADC, the randomness is reduced by at most $b_\mathrm{ADC}$ bits, with 
\begin{align}
b_\mathrm{ADC} = \log{ \sup_{f} |J_{f}|  } \, ,
\label{eq:adcdigitizationnoise}
\end{align}
where $|J_{f}|$ can be evaluated experimentally.

\subsection{Finite detection bandwidth}
\label{sec:finite_detection_bandwidth}

For a detector with linear response, the signal measured at a given time $t$ contains terms depending linearly on the fields at previous times.
Equation (\ref{HPic}) is thus replaced by
\begin{align}
y(t) = g \sqrt{\eta} \, \tilde x(t) + g \sqrt{1-\eta}\, \tilde u(t) \, ,
\end{align}
where
\begin{align}
\tilde x(t) & = \chi^x_0 \, x(t) + \sum_{t':t'<t} \chi^x_{t'} \, \tilde x(t') \, , \\
\tilde u(t) & = \chi^u_0 \, u(t) + \sum_{t':t'<t} \chi^u_{t'} \, \tilde u(t') \label{ut}
\end{align}
are stochastic processes.
The variables $x(t)$ and $u(t)$ are fresh, i.i.d.~random variables for any $t$. These are linearly mixed with past values $\tilde x(t')$ and $\tilde u(t')$ weighted by the coefficients $\chi^x_{t'}$ and $\chi^u_{t'}$.
Since the variable $u(t)$ describes the vacuum state, we have $\langle u(t)^2 \rangle =1$ and the process $u(t)$ is Gaussian.
For simplicity, we assume that the measurements take place at integer times, $t = \dots, -2, -1, 0, 1, 2, \dots$, virtually extending from $-\infty$ to $\infty$.

The min-entropy analysis that we have outlined for the i.i.d.~case can be repeated here with minor modifications. 
We obtain a result analogous to Eq.~(\ref{c9m43}), but with the factor $g$ replaced by the product $g \chi^u_0$:
\begin{align}
H_\mathrm{min}(Y|U'E)_\rho \geq 
- \log\left( \frac{1}{ 2\pi g \chi^u_0 \sqrt{1-\eta}} \right) \, ,
\end{align}
where $Y$ represents the current measurement outcome at time $t$, and $U'$ is the collection of past values of $u(t')$, for $t' < t$.
The analysis of the ADC and ADC noise yield the same formal results with the replacement $g \to g \chi^u_0$.

It remains to estimate the quantity
$g \chi^u_0$.
This goal can be accomplished offline during a calibration phase where both $\tilde x$ and $\tilde u$ are stationary Gaussian processes, and are statistically independent.
The auto-correlation function of the stochastic process $y$ is
\begin{align}
    \chi_{y,n} = \langle \tilde y(t) \tilde y(t+n) \rangle \, ,
\end{align}
which does not depend on $t$ if stationary.
From this we compute the power spectral density
\begin{align}
    f_{y}(\lambda) = \sum_{n=-\infty}^\infty \chi_{y,n} \, e^{i n \lambda}
    \, ,
\end{align}
for $\lambda \in [0,2\pi]$.
Analogously, we may compute the power spectral densities $f_{g \sqrt{\eta}\,\tilde x}$ and $f_{g \sqrt{1-\eta}\,\tilde u}$ for the re-scaled processes $g \sqrt{\eta}\,\tilde x$ and $g \sqrt{1-\eta}\, \tilde u$. As the latter are statistically independent, their power spectral densities sum up:
\begin{align}
    f_{y}(\lambda) =
    f_{g \sqrt{\eta}\,\tilde x}(\lambda) 
    + f_{g \sqrt{1-\eta}\, \tilde u}(\lambda) \, .
\end{align}
In particular, if we can experimentally estimate $f_{y}$ and $f_{g \sqrt{\eta}\,\tilde x}(\lambda)$, the power spectrum of $g \sqrt{1-\eta} \, \tilde u$ is readily computed as
\begin{align}
    f_{g \sqrt{1-\eta} \, \tilde u}(\lambda) =
    f_{y}(\lambda) -
    f_{g \sqrt{\eta} \,\tilde x}(\lambda) 
    \, .
\end{align}

Our strategy to estimate $g \chi_0^u$ relies on the notion of entropy rate of a stochastic process~\cite{Covers1991ElementsTheory}. Given a sequence of random variables $Z_1, Z_2 \dots$, the entropy rate is defined as
\begin{align}
    h[Z] = \lim_{k \to\infty} \frac{1}{k} H[Z_1 Z_2 \cdots Z_k] \, ,
\end{align}
where $H[Z_1 Z_2 \dots Z_k]$ is the joint Shannon entropy.
For a stationary process, the entropy rate can be computed from the spectral density $f_Z$ \cite{Gray2006},
\begin{align}
h[Z] 
= \frac{1}{2} \ln{(2 \pi e)} + \frac{1}{2}  \int_0^{2\pi} \frac{d\lambda}{2\pi} \ln{[f_{Z}(\lambda)]}  \, .
\end{align}
Here we are interested in computing the entropy rate of $g \sqrt{1-\eta} \, \tilde u$, which reads
\begin{align}
h[g \sqrt{1-\eta} \, \tilde u] 
& = \frac{1}{2} \ln{(2 \pi e)} + \frac{1}{2}  \int_0^{2\pi} \frac{d\lambda}{2\pi} \ln{[f_{g \sqrt{1-\eta} \,\tilde u}(\lambda)]} \, .
\label{er1}
\end{align}

Alternatively, if $Z$ is a stationary Gaussian process, the entropy rate can be computed as \cite{Covers1991ElementsTheory}
\begin{align}
    h[Z] = \lim_{k\to\infty}
    H[Z_k | Z_{k-1} Z_{k-2} \cdots] \, ,
\end{align}
which is the limit of the entropy of $Z$ conditioned on all past values.
Note that, since the variables $u(t)$ are Gaussian, the stochastic process $g \sqrt{1-\eta} \, \tilde u$ is also Gaussian. 
From Eq.~(\ref{ut}) we can readily compute the entropy of $g \sqrt{1-\eta} \, \tilde u(t)$ conditioned on past values at $t'<t$:
\begin{align}\label{er2}
h[g \sqrt{1-\eta} \, \tilde u] 
= \frac{1}{2} \ln{( 2 \pi e \sigma_{U,c}^2 )} \, , 
\end{align}
where 
\begin{align}\label{condv}
\sigma_{U,c}^2 =  g^2 (1-\eta) (\chi_0^u)^2 
\langle u(t)^2 \rangle 
= g^2 (1-\eta) (\chi_0^u)^2
\end{align}
is the conditional variance.

By equating the two expressions obtained above for the entropy rate, in Eqs.~(\ref{er1}) and (\ref{er2}), we obtain
\begin{align}
\frac{1}{2} \ln{(2 \pi e)} + \frac{1}{2}  \int_0^{2\pi} \frac{d\lambda}{2\pi} \ln{[f_{g \sqrt{1-\eta} \,\tilde u}(\lambda)]}
= \frac{1}{2} \ln{( 2 \pi e \sigma_{U,c}^2 )} \, .
\end{align}
This latter equation, together with Eq.~(\ref{condv}), allows us to compute the finite-bandwidth--gain product $g \chi^u_0$:
\begin{align}
g \chi^u_0  = \frac{1}{ \sqrt{1-\eta} } \exp{\left\{ \int_0^{2\pi} \frac{d\lambda}{4\pi} \ln{[f_{g \sqrt{1-\eta} \,\tilde u}(\lambda)]} \right\}}  \, .
\label{cond_gain_eq}
\end{align}

\subsection{Estimating out-of-bound ADC samples}
\label{sec:out_of_bound_samples}

The goal of this characterization procedure is to determine the parameter $P$ in Eq.~(\ref{eq_min_entr1}) with its confidence interval. For a total observed number of samples $N$, our best guess for $P$ is
\begin{align}
    \hat{P}={N_1\over N},
\end{align}
where $N_1$ is the number of samples falling into the range. A confidence interval can be obtained using the Hoeffding tail bound,
\begin{align}
    \mathrm{Pr} \left( \hat P - t > P  \right) < e^{- 2 N t^2} \, .
\end{align}
After a change of variables, we obtain that the bound 
\begin{align}
    \log{P} \geq \log{\left[\hat P - \sqrt{ \frac{ \ln{(1/\varepsilon_P)} } {2 N } }   \right]}
\label{eq_out_of_bound}
\end{align} 
holds true up to a failure probability not larger than $\varepsilon_P$.

\subsection{Measurement of the ADC digitization error}
\label{sec:measurement_of_ADC_digitization_error}

\begin{figure}[h]
    \centering
    \includegraphics[width=\linewidth]{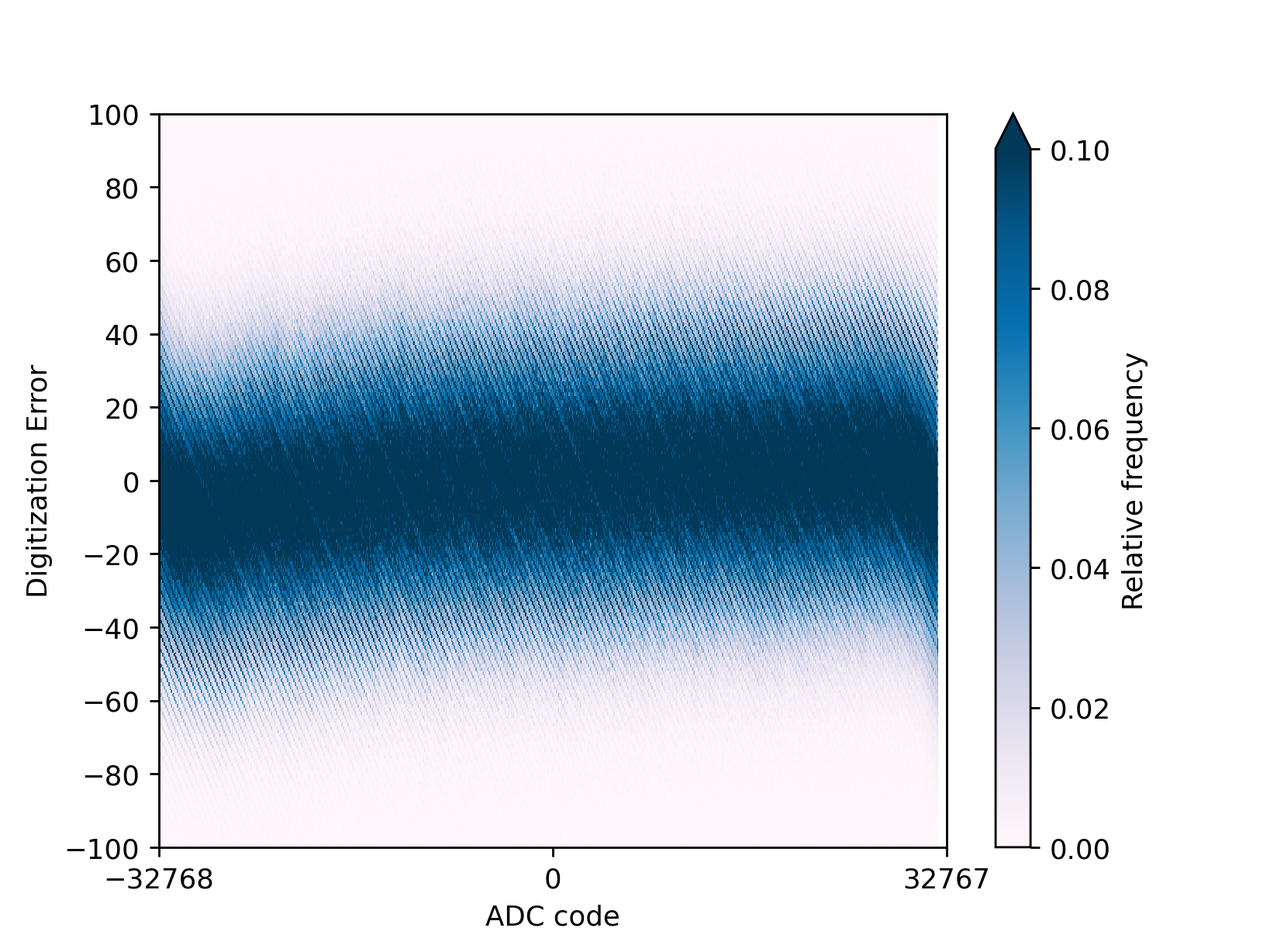}
    \caption{\textbf{Digitization error of the ADC.} The graph shows the static digitization error for each ADC code.}
    \label{fig:adcnonlinearity}
\end{figure}

Based on the method used in Ref.~\citenum{Bruynsteen2022} we measured the digitization error of our 16 bit ADC. The result is shown in Fig.~\ref{fig:adcnonlinearity}. 
However, unlike Ref.~\citenum{Bruynsteen2022} the histograms for the individual ADC codes are not Gaussian due to the interleaving nature of the ADC. Therefore we resort to Eq.~(\ref{eq:adcdigitizationnoise}) to determine the min-entropy reduction.
This yields a conservative bound as it considers the worst-case scenario which does not leverage prior information. The resulting min-entropy reduction is 7.8 bit.

\subsection{Extracting random numbers}

We implemented a real-time Toeplitz randomness extractor in FPGA. Hashing is performed by multiplying the input string of $l$ bits to a Toeplitz matrix of size $k\times l$ bits, so the output is of length $k$. The hash function is hence uniquely defined by the Toeplitz matrix, which has to be chosen randomly. This requires $k+l-1$ uniformly random bits, known as the seed. The seed should represent the output of an independent source with a statistical parameter $\varepsilon_\mathrm{seed}$, small enough for the application using the random numbers. 

The Toeplitz matrix multiplication is implemented on an FPGA using the submatrix method \cite{Zhang2016, Huang2019, Zheng2019, Gehring2021Homodyne-basedSide-information} to increase the efficiency. By splitting the seed into submatrices with 64 columns and $l$ rows, random data can be streamed into the FPGA at 64 bits and a single submatrix multiplication can be achieved each clock cycle consuming all of the data available. After $k/64$ clock cycles all of the submatrices have been applied to random numbers, and the result of each row are added together with a modulo 2 operation to provide the final result.

\section*{Author contributions}
ULA and TG conceived the idea and supervised the project.
DSN set up the experiment and tested components. TG designed and built the electronics. RZ recorded and analyzed the data under supervision of TG.
CL conceived and worked out the security proof.
TR developed the FPGA firmware. 
All authors contributed to the manuscript.

\section*{Competing interests}
ULA and TG declare competing financial interests. All other authors declare no competing interests.

\section*{Acknowledgements}
We thank Nitin Jain for helpful comments on the manuscript. DSN, RZ, TR, ULA and TG acknowledge support from the Danish National Research Foundation, Center for Macroscopic Quantum States (bigQ, DNRF142). DSN, TR, ULA and TG acknowledge support from the Innovation Foundation (grant agreement no.\ 9122-00126 and 0175-00018B). RZ and TG acknowledge funding from the Carlsberg Foundation (CF21-0466). C.L. acknowledges financial support from PNRR MUR project PE0000023-NQSTI.

\bibliography{references}

\end{document}